\documentclass{emulateapj}

\begin{document}

\title{Thermal and dynamical properties of gas accreting onto a
  supermassive black hole in an AGN}
\author{M. Mo{\'s}cibrodzka$^{1,\dagger}$, D. Proga$^{1}$ }
\affil{$^1$ Department of Physics and Astronomy, University of Nevada, 
4505 South Maryland Parkway, Las Vegas, NV 89154}
\email{$\dagger$ Current address: Department of Astrophysics/IMAPP,
Radboud University Nijmegen, P.O. Box 9010, NL - 6500 GL Nijmegen, The
Netherlands, m.moscibrodzka@astro.ru.nl}

\begin{abstract}
We study stability of gas accretion in Active Galactic Nuclei (AGN). Our grid
based simulations cover a radial range from 0.1 to 200 pc, which may enable to
link the galactic/cosmological simulations with small scale black hole
accretion models within a few hundreds of Schwarschild radii. Here, as in
previous studies by our group, we include gas radiative cooling as well as
heating by a sub-Eddington X-ray source near the central supermassive black
hole of $10^8 M_{\odot}$. Our theoretical estimates and simulations show that
for the X-ray luminosity, $L_X \sim 0.008~L_{Edd}$, the gas is thermally and
convectivelly unstable within the computational domain. In the simulations, we
observe that very tiny fluctuations in an initially smooth, spherically
symmetric, accretion flow, grow first linearly and then non-linearly.
Consequently, an initially one-phase flow relatively quickly transitions into
a two-phase/cold-hot accretion flow. For $L_X = 0.015~L_{Edd}$ or higher, the
cold clouds continue to accrete but in some regions of the hot phase, the gas
starts to move outward.  For $L_X < 0.015~L_{Edd}$, the cold phase
contribution to the total mass accretion rate only moderately dominates over
the hot phase contribution. This result might have some consequences for
cosmological simulations of the so-called AGN feedback problem.  Our
simulations confirm the previous results of Barai et al. (2012) who used
smoothed particle hydrodynamic (SPH) simulations to tackle the same
problem. However here, because we use a grid based code to solve equations in
1-D and 2-D, we are able to follow the gas dynamics at much higher spacial
resolution and for longer time in comparison to the 3-D SPH simulations.  One
of new features revealed by our simulations is that the cold condensations in
the accretion flow initially form long filaments, but at the later times,
those filaments may break into smaller clouds advected outwards within the hot
outflow. Therefore, these simulations may serve as an attractive model for the
so-called Narrow Line Region in AGN.
\end{abstract}


\section{Introduction}

Physics within the central parsecs of a galaxy is dominated by the
gravitational potential of a compact supermassive object. In a classical
theory of spherical accretion by \citet{bondi:1952}, Bondi radius $R_B$
determines the zone of the gravitational influence of a central object and it
is given by $R_{B} \approx 150 (M_{BH}/10^8 M_{\odot}) (T_{\infty}/10^5 {\rm
  K})^{-1} \, {\rm pc}$, where $M_{BH}$ is the central object mass, and
$T_{\infty}$ is the temperature of the uniform surrounding medium.  At radii
smaller than the Bondi radius, $R_{B}$, the interstellar medium (ISM) or at
least its part is expected to turn into an accretion flow.

Physics of any part of a galaxy is complex. However near the Bondi radius, it
is particularly so because there, several processes compete to dominate not
only the dynamical state of matter but also other states such as thermal and
ionization. Therefore studies of the central parsec of a galaxy require
incorporation of processes and their interactions that are typically
considered separately in specialized areas of astrophysics, e.g., the black
hole accretion, physics of ISM and of the galaxy formation and evolution. One
of the main goals of studying the central region of a galaxy is to understand
various possible connections between a supermassive black hole~(SMBH) and its
host galaxy.

Electromagnetic radiation provides one of such connections. For example, the
powerful radiation emitted by an AGN, as it propagates throughout the galaxy,
can heat up and ionize the ISM.  Subsequently, accretion could be slowed down,
stopped or turned into an outflow if the ISM become unbound.  Studies of
heated accretion flows have a long history. Examples of early and key works
include \citet{ostriker:1976}, \citet{cowie:1978}, \citet{mathews:1978},
\citet{stell:1982}, \citet{bisnovatyi:1980}, \citet{krolik:1983}, and
\citet{balbus:1989}.

The accretion flows and their related outflows are very complex phenomena.  It
is likely that several processes are responsible for driving an outflow, i.e.,
not just the energy of the radiation, as mentioned above, but also for
example, the momentum carried by the radiation.  Therefore, our group explored
combined effects of the radiation energy and momentum on the accretion flows
and on producing outflows (e.g.  \citealt{proga:2007}; \citealt{proga:2008};
\citealt{kurosawa:2008}; \citealt{kurosawa:2009a}; \citealt{kurosawa:2009b};
\citealt{kurosawa:2009c}). These papers reported on results from simulations
carried out using Eulerian finite difference codes where effects of gas
rotation and other complications such non-spherical and non-azimuthal effects
were included (see e.g. \citealt{janiuk:2008}).

To identify the key processes determining the gas properties (here, we are
mainly concerned with thermal properties) and to establish any code
limitations in modeling an accretion flow, in this paper we adopt a relatively
simple physical set up. Namely, the modeled system consists of a central SMBH
of mass $M_{BH}= 10^8 M_{\odot}$ and a spherical shell of gas inflowing to the
center. The simulations focus on regions between 0.1-200 pc from the central
object, where the outer boundary is outside of $R_B$. The key difference
compared to the Bondi problem is an assumption that the central accretion flow
is a point-like X-ray source. The X-rays illuminate the accreting gas and the
gas itself is allowed to cool radiatively under optically thin conditions.  To
keep the problem as simple as possible, the radiation luminosity is kept fixed
instead of being computed based on the actual accretion rate for an assumed
radiation efficiency (see also \citealt{kurosawa:2009a}).
 
To model the presented problem, one needs to introduce extra terms into the
energy equation to account for energy losses and gains. The physics of an
optically thin gas that is radiatively heated and cooled, in particular, its
thermal and dynamical stability has been analyzed in a great detail by
\citet{field:1965}. Therefore, to study thermal properties of accretion flows
or dynamical properties of thermally unstable gas, it is worthwhile to combine
\citet{bondi:1952} and \citet{field:1965} theories. Notice, that our set-up is
very similar that that used in the early works in the 70-ties and 80-ties that
we mentioned above. Some kind of complexity and time variability in a heated
accretion flow is expected based on the 1-D results from the early work.

A dynamical study of the introduced physical problem requires resolving many
orders of magnitude of the radial distance from the black hole.  Our goal is
not only to cover the largest radial span as possible but also to resolve any
small scale structure of the infalling gas.  This is a challenging goal.  To
study the dynamics of gas in a relatively well controlled computer experiment,
we use the Eulerian finite difference code ZEUS-MP \citep{hayes:2006}.

We address systematically numerical requirements to adequately treat the
problem of thermally unstable accretion flows.  We introduce an accurate
heating-cooling scheme that incorporates all relevant physical processes of
X-ray heating and radiative cooling.  Low optical thickness is assumed, which
decouples fluid and radiation evolution.  We resolve three orders of magnitude
in the radial range by using a logarithmic grid where the logarithm base is
adjusted to the physical conditions. We solve hydrodynamical equations in 1-
and 2- spatial dimensions (1-D and 2-D).  We follow the flow dynamics for a
long time scale in order to investigate the non-linear phase of gas
evolution. Notice that most of the earlier work in the 70-ties and 80-ties,
focused on linear analysis of stability, early stages of evolution of the
solutions, and considered only 1-D cases.

As useful our group's past studies are, we keep in mind that any result should
be confirmed by using more than one technique or approach. Therefore,
\citet{barai:2012} (see also \citealt{barai:2011}), began a parallel effort to
model accretion flows including the same physics but instead of performing
simulations using a grid based code we used the smoothed particle hydrodynamic
(SPH) GADGET-3 code \citet{springel:2005}.  Overall, the 3-D SPH simulations
presented by \citet{barai:2012} showed that despite this very simple set up,
accretion flows heated by even a relatively weak X-ray source (i.e., with the
luminosity around 1\% of the Eddington luminosity) can undergo a complex time
evolution and can have a very complex structure.  However, the exact nature
and robustness of these new 3-D results has not been fully established.
\citet{barai:2012} mentioned some numerical issues, in particular, artificial
viscosity and relatively poor spacial resolution in SPH, because of the usage
of linear length scale (as opposed to logarithmic grid in ZEUS-MP), limit
ability to perform a stability analysis where one wishes to introduce
perturbations to an initially smooth, time independent solution with well
controlled amplitude and spatial distribution (SPH simulations have intrinsic
limitations in realizing a smooth flow). Therefore, the robustness and
stability of the solutions found in the SPH simulations are hard to access due
to mixing of physical processes and numerical effects.  Here, we aim at
clarifying the physics of these flows and measure the role of numerical
effects in altering the effects of physical processes.

Our ultimate goal is to provide insights that could help to interpret
observations of AGN. We explore the conditions under which the two phase, hot
and cold, medium near an AGN can form and exist. Such two phase accretion
flows can be a hint to explain the modes of accretion observed in galactic
nuclei, but also to explain the formation of broad and narrow lines which
define AGN.  We also measure the so-called covering and filling factors and
other quantities in our simulations in order to relate the simulation to the
origin of the broad and narrow line regions (BLRs and NLRs, respectively).
The connection of this work to the galaxy evolution and cosmology is that we
resolve lower spatial scales, and hence can probe what physical processes
affect the accretion flow.  In our models, we can directly observe where the
hot phase of accretion turns into a cold one or where an eventual outflow is
launched. In most of the current simulations of galaxies
(e.g. \citealt{dimatteo:2012} and references therein) these processes are
assumed or modeled by simple, so called sub-resolution, approximations
because, contrary to our simulations, the resolution is too low to capture the
flow properties on adequately small scales.
 
The article is organized as follows. In \S~\ref{sec:equation}, we present the
basic equations describing the physical problem.  In \S~\ref{sec:num_setup},
we show the details of the numerical set up. Results are in
\S~\ref{sec:results_1d} and in \S~\ref{sec:results_2d}.  We summarize the
results in \S~\ref{sec:discussion}.

\section{Basic Equations}\label{sec:equation}

We solve equations of hydrodynamics:
\begin{equation}
\frac{D\rho}{Dt} + \rho {\bf \nabla \cdot v}=0 \label{eq:mass}
\end{equation}
\begin{equation}
\rho \frac{D{\bf v}}{Dt} = -\nabla P + \rho {\bf g} \label{eq:mom}
\end{equation}
\begin{equation}
\rho \frac{D}{Dt} (\frac{e}{\rho})= -P {\bf \nabla \cdot v} + \rho
     {\mathcal L} \label{eq:energy}
\end{equation}
where $D/Dt$ is Lagrangian derivative and all other symbols have their
usual meaning. To close the system of equations we adopt the
$P=(\gamma-1)e$ equation of state where $\gamma = 5/3$. Here $g$ is the
gravitational acceleration near a point mass object in the center. The
equation for the internal energy evolution has an additional term
$\rho {\mathcal L}$, which accounts for gas heating and
cooling by continuum X-ray radiation produced by an accretion flow
near the central SMBH.
The heating/cooling function contains four terms which are: (1)
Compton heating/cooling ($G_{Compton}$), (2) heating and cooling due
to photoionization and recombination ($G_X$), (3) free-free
transitions cooling ($L_{b}$) and (4) cooling via line emission
($L_{l}$) and it is given by (\citealt{blondin:1994}, \citealt{proga:2000}):
\begin{equation}
\rho  {\mathcal L} =   n^2 (G_{Compton} + G_X - L_b - L_l ) \, \, {\rm [erg
\,\, cm^{-3} s^{-1} ]}\label{eq:HC_full}
\end{equation}
where
\begin{equation}
G_{Compton}=\frac{k_b \sigma_{TH}}{4 \pi m_e c^2} \xi T_X \left(1- \frac{4T}{T_X}\right)
\end{equation}
\begin{equation}\label{eq:HC_2}
G_X= 1.5 \times 10^{-21} \xi^{1/4} T^{-1/2} \left(1-\frac{T}{T_X}\right) 
\end{equation}
\begin{equation}
L_b=  \frac{2^5 \pi e^6}{ \sqrt{27} h m_e c^2 } \sqrt{\frac{2\pi k_b T}{m_ec^2}}
\end{equation}
\begin{equation}\label{eq:HC_4}
L_l=   1.7 \times 10^{-18} \exp\left(- \frac{1.3 \times 10^5}{T}\right) \xi^{-1} T^{-1/2} -
10^{-24} 
\end{equation}
where $T_X$ is the radiative temperature of X-rays and $T$ is the
temperature of gas. We adopt a
constant value $T_X = 1.16 \times 10^8$ K ($E=10$ keV) at all times.
The numerical constants in Equation~\ref{eq:HC_2} and~\ref{eq:HC_4} are taken from an
analytical formula fit to the results from a photoionization code
XSTAR \citep{kallman:2001}. XSTAR calculates the ionization structure and
cooling rates of a gas illuminated by X-ray radiation using atomic data.
The photoionization parameter $\xi$ is defined as:
\begin{equation}
\xi \equiv  \frac{4 \pi F_X}{n} = \frac{L_X} {n r^2}  = \frac{f_X L_{Edd}}
    {n r^2} = \frac{f_X L_{Edd} m_p \mu} {\rho r^2}  \,\, {\rm [ergs \,\, cm
        \,\, s^{-1}]} 
\end{equation}
where $F_X$ is the radiation flux, $n=\rho/(\mu m_p)$ is the number
density, and $\mu$ is a mean molecular weight. Given $\xi$ definition,
notice that $\mathcal L$ is a function of thermodynamic variables but
also strongly depends on the distance from the SMBH. 

The luminosity of the
central source $L_X$ is expressed in units of the Eddington
luminosities, $f_X \equiv L_X/L_{Edd}$. The reference Eddington luminosity for a
supermassive black hole mass considered in this work is
\begin{equation}
L_{Edd} \equiv \frac{ 4 \pi G M_{BH} m_p c}{\sigma_{TH}} = 1.25 \times 10^{46}
\left(\frac{M }{10^8 M_{\odot}}\right) \,\,
{\rm [ergs \,\, s^{-1}]}
\end{equation}

\section{Method and Initial Setup}\label{sec:num_setup}

To solve Equations.~\ref{eq:mass}, \ref{eq:mom}, \ref{eq:energy},
we use the numerical code ZEUS-MP \citep{hayes:2006}. We modify the original
version of the code in particular, we use a Newton-Raphson method to
find roots of Equation~\ref{eq:energy} numerically at each time
step. We have successfully tested the numerical method against an
analytical model with heating and cooling. We describe the numerical
code tests in the Appendix, showing the thermal instability (TI)
development in the uniform medium.

We solve equations in spherical-polar coordinates. Our computational domain
extends in radius from 0.1 to 200 pc. The useful reference unit is a radius of
the innermost stable circular orbit of a central black hole: $r_*= 6
GM_{BH}/c^2$. We assume the fiducial mass of the black hole $M_{BH}=10^8
M_{\odot}$ for which $r_*=8.84 \times 10^{13} {\rm cm}$. The computational
domain in these units ranges from $r_i=3484.2 r_*$ to $r_o=6.9683 \times 10^6
r_*$ (or $r_i = 6.6 10^{-4} R_B$ and $r_o=1.3 R_B$, where $R_B = 152$
pc). Since $r_i$ is relatively large in comparison to the BH horizon we cannot
model here the compact regions near the black hole where X-ray emission is
produced. Instead we parameterize the X-ray luminosity using $f_X$, so that
$L_X=f_X L_{Edd}$.  We solve equations for five values of $f_X$=0.0005, 0.008,
0.01, 0.015, 0.02 (these numbers correspond to models later labeled as A, B,
C, D and E).

As initial conditions for the A model (lowest luminosity), we use an
adiabatic, semi-analytical solution from \citet{bondi:1952}. For higher
luminosities the integration of equations starts from last data from a model
with one level lower luminosity provided that the lower $f_X$ solution is
time-independent.  The procedure is adopted in order to increase the
luminosity in a gradual manner rather than sudden.~\footnote{ We also decouple
  $L_X$ from $\dot{M}$ in order to avoid introducing additional parameters
  into the equations.  While coupling these quantities not only a radiative
  efficiency of gravitational to radiative energy has to be assumed but one
  also needs to know how to calculate the mass accretion rate at the very
  compact region way below $r_i=0.1 pc$. Another reason for decoupling $L_X$
  and $\dot{M}$ is that we are interested in caring out a stability analysis
  and perturb a steady state solutions with all model parameters fixed.}  Only
for steady state solutions (with assumption that the mass accretion rate is
constant from $r_i$ to $r_*$) the efficiency of conversion of gravitational
energy into radiation $\eta$ is related to $f_X$ as
\begin{equation}
\frac{\eta}{\eta_r}  = \frac{f_X}{ \dot{m}} 
\end{equation}
where $\dot{m}$ is a mass accretion rate in Eddington units
($\dot{M}_{Edd}=L_{Edd}/\eta_r c^2$ and $\eta_r=0.1$ is a reference
efficiency) and it is measured from the model data.  In our steady
state models, $\dot{m}
\approx 1$, therefore the energy conversion efficiency
in these cases is approximately $\eta=0.1f_X$. 

Our boundary conditions put constrains on a density at $r_o$, it is
set to be $\rho_o=10^{-23} {\rm g \, cm^{-3}}$. For other variables we
use an outflow type of boundary conditions at the inner and outer
radial boundary. In 2-D models our computational domain extends in
$\theta \in (0,90^\circ)$. At the symmetry axis and at the equator we use
appropriate reflection boundary conditions.
The numerical resolution used depends on the number of dimensions
i.e.: in 1-D $N_r=256, 512, 1024, 2048, 4096$; in 2-D
$(N_r,N_\theta)=(256,64), (512,128), (1024,256)$. The spacing of the
radial grid is set as $dr_i/dr_{i+1}$=1.023, 1.01, 1.0048, 1.002,
1.0008 for $N_r$=256, 512, 1024, 2048, and 4096, respectively. The
number of grid points in the second dimension are chosen so
that the linear size of the grid zone in all directions is similar (i.e., $r_i
\Delta \theta_j \approx \Delta r_i$).

\section{Results: 1-D models}\label{sec:results_1d}

\subsection{1-D Steady Solutions}

We begin with presenting the basic characteristics of 1-D solutions.
Table~\ref{tab:1d} shows a list of all our 1-D simulations. 
Each simulation was performed until $t_{f}=20$ Myr equivalent to
4.7 dynamical time scales at the outer boundary $r_o=200$pc
($t_{dyn}=t_{ff}=\sqrt{r_o^3/2GM_{BH}}=4.21$ Myr). Only some of the
numerical solutions settled down to a time-independent state at $t_{f}$. 
We focus on analyzing two representative solutions, that are steady-state at 
$t_{f}$: 1D256C and 1D256D, with the X-ray luminosity of the former
$f_X=10^{-2}$, and of the latter $f_X=1.5 \times 10^{-2}$.
Note that these solutions were obtained using the lowest resolution.
We find these two solutions instructive in showing the thermal
 properties of the gas.
\begin{table*}[!ht]
\begin{center}
\begin{tabular}{ccccccccc}
  \hline
  Model ID & $f_X$ & $N_r$ & $t_f$ & $\langle\dot{M}\rangle_t$& $\langle\chi\rangle_{r,t}$&
 $\langle\tau_{X,sc}\rangle_t$& Max($\tau_{X,sc}$)&comment \\
  & & & [Myr] & ${\rm [M_{\odot} \, yr^{-1}]}$& && \\
  \hline
  1D256A  & $5 \times 10^{-4}$ & 256  & 20 &2.0 &3.9&0.44& 0.49& s\\
  1D512A  & $5 \times 10^{-4}$ & 512  & 20&2.0 &6.1&0.45& 0.51 & s\\
  1D1024A & $5 \times 10^{-4}$ & 1024 & 20&2.0 &6.8&0.46& 0.52 & s\\
  1D2048A & $5 \times 10^{-4}$ & 2048 & 20&2.0 &6.8&0.46& 0.53 & s\\
  1D4096A & $5 \times 10^{-4}$ & 4096 & 20&2.0 &6.8&0.46& 0.53 & s\\
 &&\\
  1D256B  & $8 \times 10^{-3}$ & 256  & 20&1.8 &0   & 0.1 & 0.12&s\\
  1D512B  & $8 \times 10^{-3}$ & 512  & 20&1.8 &0   & 0.1 & 0.13&s\\
  1D1024B & $8 \times 10^{-3}$ & 1024 & 20&1.8 &0   & 0.1 & 0.13&s\\
  1D2048B & $8 \times 10^{-3}$ & 2048 & 20&1.9 &0   & 0.1 & 1.7 &s\\
  1D4096B & $8 \times 10^{-3}$ & 4096 & 20&1.95&0.05& 0.13& 5.8 &ns\\
&&\\
  1D256C  & $1 \times 10^{-2}$ & 256  & 20&1.7 & 0    & 0.09 & 0.09 &s\\
  1D512C  & $1 \times 10^{-2}$ & 512  & 20&1.8 & 0    & 0.09 & 0.09 &s\\
  1D1024C & $1 \times 10^{-2}$ & 1024 & 20&1.8 & 0    & 0.11 & 10.4 &s\\
  1D2048C & $1 \times 10^{-2}$ & 2048 & 20&1.8 & 0.11 & 0.13 & 9.3  &ns\\
&&\\
  1D256D  & $1.5 \times 10^{-2}$ & 256  &20&1.5 & 0.7 & 0.2  & 24 & s\\
  1D512D  & $1.5 \times 10^{-2}$ & 512  &20&1.5 & 1.9 & 0.23 & 28 & ns\\
  1D1024D & $1.5 \times 10^{-2}$ & 1024 &20&2.1 & 5.8 & 0.55 & 61 & ns\\
&&\\
  1D256E  & $2 \times 10^{-2}$ & 256 & 20 &1.23 & 4.1 & 0.2 & 30 &ns\\
\hline
\end{tabular}
\caption{
List of 1-D solutions. Columns from left to right are: model ID, $f_X$ -
dimensionless luminosity, $N_r$ - numerical resolution in radial direction,
$t_f$ - final time of the simulation, $\langle\dot{M}\rangle_{t}$ - time
averaged mass accretion rate, $\chi$ - a ratio of cold to hot mass accretion
rate averaged over radii less than $10^{20}$ cm and simulation time, (cold
phase is any gas with temperature $T<10^5$ K),
$\langle\tau_{X,sc}\rangle_{\theta,t}$ - angle and time averaged optical depth,
Max($\tau_{X,sc}$) - maximum optical depth recorded during the simulation, and
comments (s-reached a steady state solution at $t_f$, ns-non steady state
solution at $t_f$).}\label{tab:1d}
\end{center}
\end{table*}

\begin{figure}
\includegraphics[angle=0,scale=0.2]{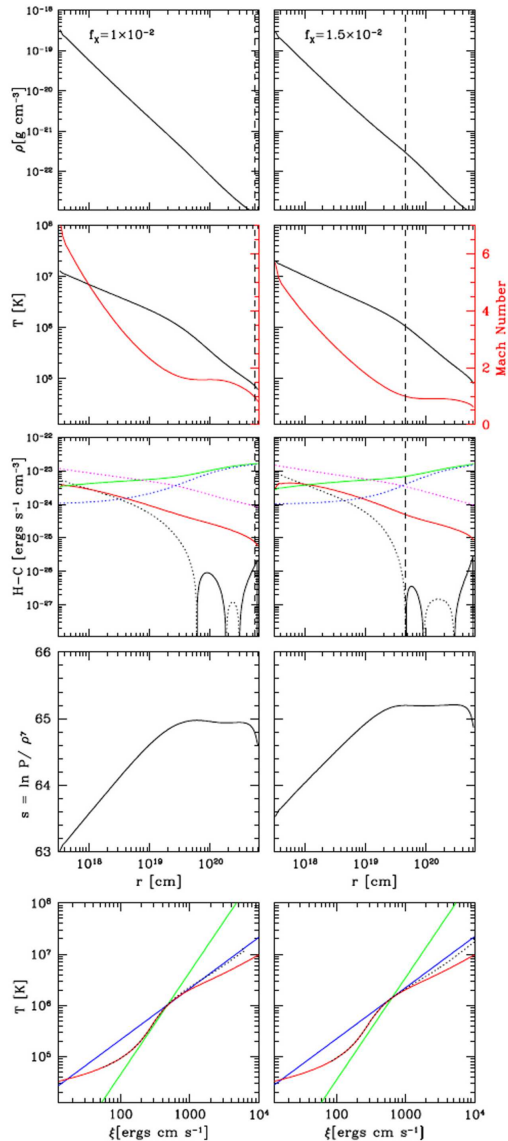}
\caption{
Structure of 1-D accretion flow in run 1D256C (left column,
$f_X=1\times10^{-2}$) and 1D256D (right column,
$f_X=1.5\times10^{-2}$). Each panel is a snapshot taken at t=20 Myr.
Panels from top to bottom show: density, temperature with Mach number
(Mach number scale is on the right hand side), heating/cooling rates,
and entropy S. The dashed vertical line in top panels marks the
position of the sonic point. In panel with heating/cooling rates the
black solid line is a net heating/cooling and color lines indicate
particular physical process included in the calculations: Compton
heating (red line), photoionization heating (green line),
bremsstrahlung cooling (magenta-line), cooling through line emission
(blue-line). The bottom panels display the gas temperature as a
function of photoionization parameter; color lines indicate gas in
radiative equilibrium (red), constant pressure conditions (blue) and
free-fall compression (green).}
\label{fig:st1d}
\end{figure}

Figure~\ref{fig:st1d} presents the overall structure of model
$1D256~C$ and $D$ (model C and D in the left and right column,
respectively). Panels from top to bottom in Figure~\ref{fig:st1d}
display: radial profiles of gas density, gas temperature overplotted
with the Mach number (red line with the labels on the right
hand side of the panels), the net heating/cooling rate plotted together with
contribution from each physical process (see Equation~\ref{eq:HC_full}), 
the entropy S, and the bottom row shows gas
temperature as a function of $\xi$. In the bottom panels, the red line
indicates the T-$\xi$ relation for radiative equilibrium 
(i.e. solving ${\mathcal L}(\xi,T)=0$ for each
$T$). The green line indicates a $T-\xi$ relation for a
gas being adiabatically compressed due to the geometry of the
spherical accretion ($T \propto \xi^2$), while the blue line
 for a constant pressure gas ($T \propto \xi^1$).

The 1D256C and D solutions differ mainly in the position of the sonic point
and in the fact that the model 1D256D  is strongly time dependent for 
a short period of time during initial evolution (see below). 
However, in most part the solution share several common properties.
In particular, in both solutions, 
the gas is nearly in radiative equilibrium at large radii
whereas, at small radii 
(below $r\approx2\times10^{19}$cm, where $T>2 \times10^6$K)
they depart from  the equilibrium quite  significantly.
In the inner and supersonic parts of the solutions, $T$ scales with 
$\xi$  as if gas was under constant pressure. 
At large radii where the solutions are nearly in the radiative equilibrium,
the net heating/cooling is not exactly zero. One can identify,
four zones where either cooling or heating dominates. 
In the most inner regions where the gas is supersonic, adiabatic
heating is very strong and the dominant radiative process is cooling 
by free-free emission. At the outer
radii, the cooling in lines and heating by photoionization
dominates. For models considered in this paper the Compton cooling is
the least important. 

Inspecting the bottom panels in Figure~\ref{fig:st1d}, one can
suspect the gas is in the middle section of the computational domain to be
thermally unstable because the slope of the $T-\xi$ relation (in the log-log
scale) is larger than 1.  Notice also that in both solutions the entropy is a
non-monotonic function of radius. The regions where the entropy decreases with
increasing radius correspond to the regions where there is net heating and the
Schwarzschild criterion indicates convective instability at these radii. We therefore
conclude that both solutions could be unstable.  We first check more
formally the thermal stability of our solutions.

\subsection{Thermal Stability of Steady Accretion Flows}

The linear analysis of the growth of thermal modes under the radiative
equilibrium conditions (${\mathcal L}(\rho_0,T_0) = 0$) has been examined in
detail by \citet{field:1965} (see Appendix for basic definitions).  In
Figure~\ref{fig:tibv}, in the top panels (left and right column correspond
again to model 1D256C and D), we show the radial profiles of various mode
timescales. The timescales, $\tau=1/n$, are calculated using
definitions~\ref{eq:Np} and~\ref{eq:Nv}.  The growth timescale of short
wavelength, isobaric condensations $\tau_{TI}=-1/N_p$ is positive (thermally
unstable zone marked with the dotted line) in a limited radial range between
about 10 and 100 pc. The location of the thermally unstable zone depends on
the central source luminosity, and it moves outward with increasing $f_X$. The
long wavelength, isochoric perturbations are damped, at all radii, on
timescales of $\tau_{v}=-1/N_v$ (faster than TI development).  The short
wavelength nearly adiabatic, acoustic waves are damped as well, and
$\tau_{ac}=-2/(N_v-N_p)$.  In Figure~\ref{fig:tibv}, the dashed line is the
accretion timescale $\tau_{acc}=r/v$.  Within the thermally unstable zone,
$\tau_{TI}$ is short in comparison to $\tau_{acc}$, in both models.

\begin{figure}
\includegraphics[angle=0,scale=0.17]{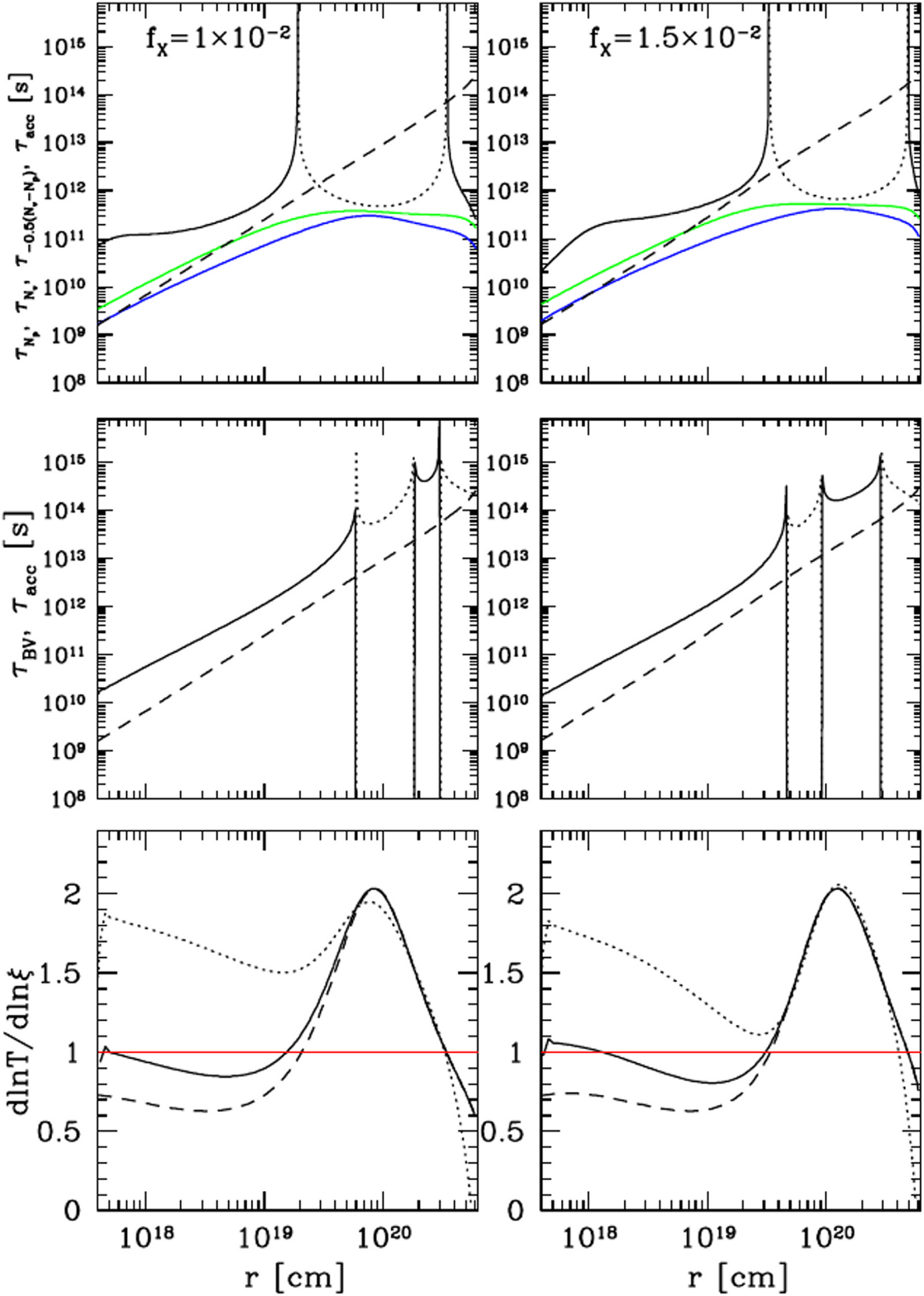}
\caption{
Left and right panels correspond, respectively to runs 1D256 C and D. Top
panels show the instability growth rates in comparison to the accretion time
scale ($\tau_{acc}=r/v$, dashed line). The time scale for the short wavelength
isobaric mode growth is displayed as the dotted line while the damping rate as
solid line ($\tau_{N_p}$).  Other two lines show the long-wavelength isochoric
mode damping rate $\tau_{N_v}$ (heavy line) and the effective acoustic waves
damping time scale $\tau_{N_v}$ (light line). Middle panel: The dashed line is
$\tau_{acc}$ and solid line is $t_{BV}=1/\omega_{BV}$, where $\omega_{BV}^2>0$
is the ${\rm Brunt-V\ddot{a}is\ddot{a}l\ddot{a}}$ oscillation frequency for a
spherical system. Solid lines show the regions which are unstable
convectivelly. The dotted line indicates region where $\omega_{BV}^2<0$ and
oscillations are possible. Bottom panels: the derivative $d \ln T/ d \ln \xi$
as a function of radius is shown as a solid line.  $(d \ln T/ d \ln \xi)_{ad}$
for an adiabatic inflow is marked as dotted line, and dashed line is the same
derivative for radiative equilibrium conditions. Horizontal one indicates
slope of~1.
\label{fig:tibv}}
\end{figure}

\citet{balbus:1986}, \citet{balbus:1989}, \citet{mathews:1978}, (and also \citealt{krolik:1983})
extended the analysis by \citet{field:1965} to
spherical systems with gravity, in more general case when initially the gas is
not in the radiative equilibrium. Their approximate solution gives the formula
for linear evolution of the short wavelength, isobaric, radial perturbation
as it moves with smooth background accretion flow
(Equation 23 in \citealt{balbus:1986} or Equation 4.12 in
\citealt{balbus:1989}). Since the two presented solutions are close to
radiative equilibrium, the approximate formula for the
growth of a comoving perturbation given by \citet{balbus:1989} reduces to
\begin{equation}
\delta (r) = \frac{\delta \rho}{\rho} =\delta_s \exp \left( \int_{r_s}^{r_f} - \frac{N_p(r')}{v(r')} dr'
\right), \label{eq:amp}
\end{equation}
where $N_p(r')$ is a locally computed growth rate of a short wavelength, isobaric
perturbation as defined in the Appendix or \citet{field:1965}, $r'$ is radius
where $N_p(r') < 0$, and $\delta_s$ is an initial amplitude of a perturbation at some
starting radius $r=r_s$. Using Equation~\ref{eq:amp}, the isobaric
perturbation amplification factors are $\delta/\delta_s \approx 10^{10},
10^{16}, 10^{19}$ and $10^{33}$, for models 1D256A, B, C, and D,
respectively. Notice that these amplification factors are calculated
for the asymptotic, maximum physically allowed growth rate, $n=-N_p$,
which might not be numerically resolved.

\begin{figure}[ht]
\includegraphics[angle=0,scale=0.38]{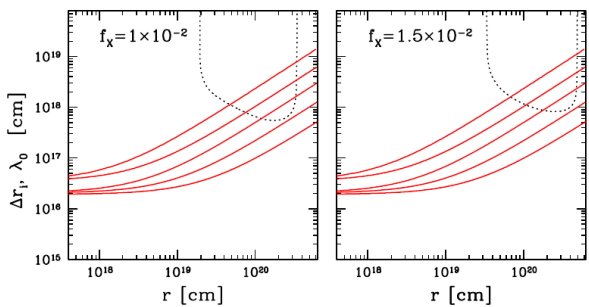}
\caption{Grid spacing (red lines) in models with $N_r$=256, 512, 1024,
  2048, and 4096 points and $\lambda_0$ (black, dotted line) as a function of radius in
  models 1D256 C (left panel) and D (right panel).
\label{fig:res}}
\end{figure}

To quantify the role of TI in our simulations we ought to address the
following question. What is the minimum amplitude and wavelength of a
perturbation in our computer models?  The smallest amplitude variability is
due to machine precision errors, $\epsilon_{machine} \approx 10^{-15}$ (for a
double float computations).  The typical $\lambda$ of these numerical
fluctuations are of the order of the numerical resolution, $\Delta r_i$.  The
discretization of the computational domain affects the TI growth rates in our
models in two ways: (1) the numerical grid refinement limits the size of the
smallest fragmentation that can be captured; (2) the rate at which the
condensation grows in the numerical simulations depends on number of points
resolving a condensation. As shown in Appendix the perturbation of a given
$\lambda$ has to be resolved by 20, or more, grid points. A wavelength
$\lambda_0$ for which $n = - 0.9973 \times N_p$ is shown in
Figure~\ref{fig:res} together with $\Delta r_i $ as a function of radius for
models with $N_r$=256, 512, 1024, 2048, and, 4096 grid points. In low
resolution models we marginally resolve $\lambda_0$.  We therefore expect the
TI fragmentations to grow slower than theoretical estimates.  Reduction of the
growth rate due to these numerical effects even by a factor of a few is enough
to suppress variability because of the strong exponential dependence.

\begin{figure}[ht]
\includegraphics[angle=0,scale=0.4]{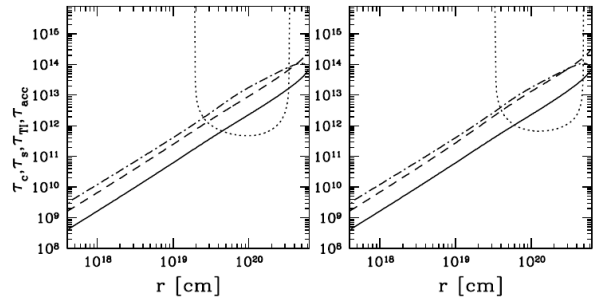}
\caption{Time scales in 1-D, stationary models
1D256 C (left panel) and 1D256 D (right panel): accretion time scale 
($\tau_{acc}$, dashed line), compression time scale ($\tau_c$, solid
  line), tidal stretching time scale ($\tau_s$, dotted-dashed line),
and condensation growth time scale ($\tau_{TI}$, dotted line). 
\label{fig:time_scales}}
\end{figure}

Thermal mode evolution depends not only on the numerical effects but also
other processes affecting the flow.  Figure~\ref{fig:time_scales} shows the
comparison of time scales of physical processes involved: the compression due
to geometry of the inflow and stretching due to accretion dynamics.  We expect
that any eventual condensation formed from the smooth background which leaves
the thermally unstable zone, would accrete with supersonic background
velocity. From the continuity equation, the co-moving density evolution is a
balance of two terms $(1/\rho) (D\rho/Dt)=-2v/r - \partial v/\partial r$,
i.e., the compression and tidal stretching. The amplitude of condensation
grows in regions where there is compression due to geometry and decreases in
regions where fluid undergoes acceleration - it stretches the perturbation. In
the models 1D256D and C interior of the TI zone, the evolution of the
perturbation is dominated by compression because the compression time scale is
the shortest.

\subsection{Convective Stability of Steady Accretion Flows}

In this subsection, we examine in more detail convective stability of 
our solutions.
In Figure~\ref{fig:tibv}, (middle panels), we compare the accretion time scale
$\tau_{acc}$ and the ${\rm Brunt-V\ddot{a}is\ddot{a}l\ddot{a}}$ time scale
 $\tau_{BV}=\frac{1}{\omega_{BV}}$ associated
with the development of convection. The frequency $\omega_{BV}$ is
defined as $\omega_{BV}^2 \equiv (-\frac{1}{\rho} \frac{\partial
P}{\partial r}) \frac{\partial \ln S}{dr}$. The convectivelly unstable
regions are marked as solid lines ($\omega_{BV}^2 >0$).
The convectivelly unstable zones 
overlap with the thermally unstable zones. 
Since $\tau_{acc} \ll \tau_{BV}$ convective motions
might not develop, at least at the linear stage of the development of TI.

In the bottom panels of Figure~\ref{fig:tibv} we show 
the logarithmic derivatives of $d \ln T / d \ln \xi$ that could
be used to graphically assess the stability of the flow.
This can be done 
by comparing the derivatives (the slopes of the 
$ln T - \ln \xi$ relation) for three cases: model data (solid
line where $T$ and $\xi$ are taken directly from the simulations),
purely adiabatic inflow (dotted line, assuming that the velocity
profile is same as in the numerical solution), and radiative equilibrium
conditions (dashed line). In particular, the regions where the solid line
is above the red line correspond to the potentially TI zones.
The regions where the dotted line is below the solid line
correspond with the zone where the flow is potentially convectivelly unstable.
The conclusion regarding the flow stability is consistent
with the conclusion reached by analyzing the time scales
shown in the top and middle panels of Fig. 2.

\subsection{Other Physical Consequences of Radiative Heating and
  Cooling - obscuration effects} 

\begin{figure}[ht]
\includegraphics[angle=0,scale=0.4]{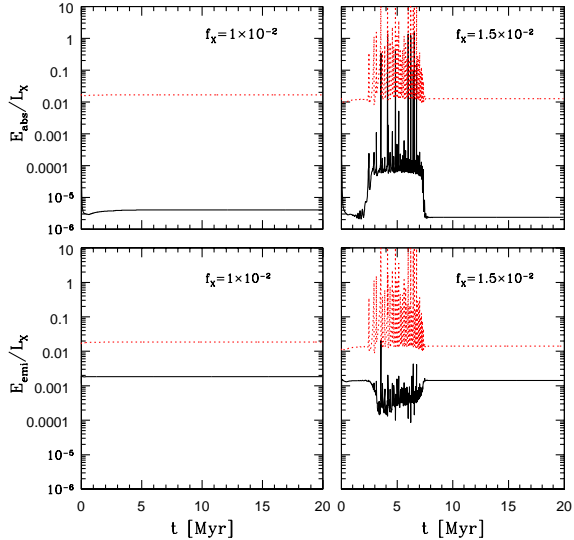}
\caption{
Fraction of central illuminating source radiative energy intrinsically absorbed (upper
panels) and emitted (bottom panels) by gas per second as a function of
time. We show the steady state solutions $1D256C$ and $1D256D$ in left and right
panels, respectively. Solid lines
show the net absorption/emission and dashed lines indicate the intrinsic
absorption and emission.
\label{fig:en1d}}
\end{figure}

The growth of the thermal instability leads to the
development of a dense cold clouds (shells in 1-D models; e.g. variable phase
in model 1D256D).  The
enhanced absorption in the dense condensations may make them optically
thick. Here we check if the
time-dependent models are self-consistent with our optically thin
assumption. Figure~\ref{fig:en1d} shows the amount of energy absorbed
and emitted by the gas (heating and cooling rates 
integrated over a volume at each time moment)
in comparison to the luminosity of the central source in models 1D256C
and D. Solid lines show the net rate of the energy exchange
between radiation and matter (cooling function ${\mathcal L}$ integrated over the
simulation volume) while dashed lines
indicate the intrinsic absorption and emission (heating and cooling terms used in
${\mathcal L}$ are integrated independently). The net heating-cooling rate is
mostly much lower than unity reflecting the fact that  
in the steady state the gas is nearly in a radiative equilibrium.  
During a variable phase (part of model 1-D256D) the energy absorbed by 
the accretion flow (black solid line) becomes comparable to 
the X-ray luminosity of the central black hole.  During this variable phase 
the optical thickness
of accreting shells can increase up to $\tau_{X,sc} \approx 20$ where
the majority contribution to opacity is due to 
photoionization absorption. The average optical thickness increases in
models with higher resolution indicating that the flow is more
variable and condensations are denser. This increase in optical depth
is related to shells condensating  much 
faster in runs with higher resolutions. The
dense condensations falling towards the center could reduce the radiation
flux in the accretion flow at larger distances.  It is beyond the scope of the
present paper to investigate the dependence of the flow dynamics on
the optical thickness effects and we leave it to the
future study. 

Significant X-ray absorption is related also to transfer of momentum
from radiation field to the gas. To estimate the importance of the momentum
exchange between radiation and matter, one can compute a
relative radiation force: 
\begin{equation}
f_{force} \equiv \frac{\sigma_{sc}+\sigma_{X}}{\sigma_{TH}} f_X  \label{eq:rforce}
\end{equation}
where $\sigma_X$ is the energy averaged X-ray cross-section. 
The momentum transfer is significant when $f_{force}>1$.
Using our expression for the heating function due to X-ray photoionization 
$\sigma_X/\sigma_{TH}=H_X / n / F_X = 2.85\times 10^4 \xi^{-3/4}
T^{-1/2}$ (see  \S~\ref{sec:equation}).
Even for a dense cold shell $f_{force}$ is at most 0.1 (in case when
$\tau_{max}\approx60$). Therefore radiation force is not likely
to directly launch an outflow. However, the situation may change when optical effects
are taken into account. 

\subsection{$\dot{M}$ Evolution}

\begin{figure}[hb]
\includegraphics[angle=0,scale=0.4]{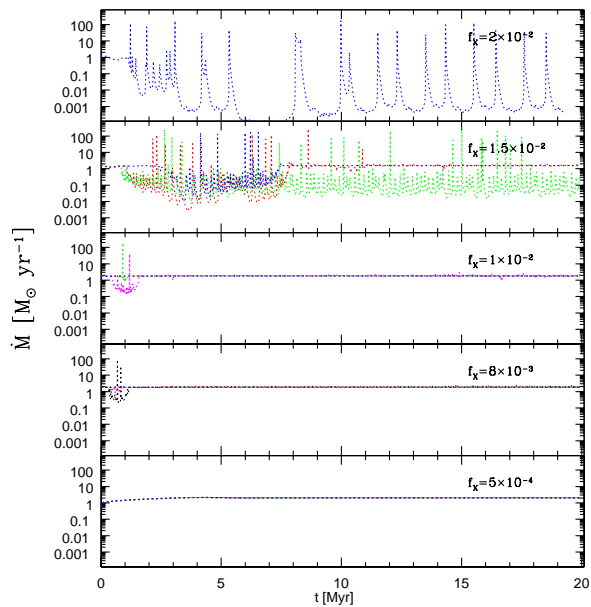}
\caption{
Mass accretion rate in 1-D solutions for $f_X=0.0005, 0.08, 0.01$, and
$0.015$. Different colors shows $\dot{M}$ for various number of grid
points: $N_r$=256 (blue), 512 (red), 1024 (green), 2048 (magenta),
4096 (black).~\label{fig:mdot1d}}
\end{figure}

We end our presentation of 1-D results with a few comments on
the time evolution of the mass accretion rate, $\dot{M}$.
Figure~\ref{fig:mdot1d} displays 
$\dot{M}$ vs time  measured for all of our 1-D models. One can divide
the solutions into two subcategories: steady and unsteady where
$\dot{M}$ varies from small fluctuations to large changes. For a given
$f_X$, the time behavior of the solution depends on the resolution,
due to effects described above. In the variable
models a fraction of the accretion proceeds in a form of a cold phase
defined as all gas with $T < 10^5$ K.  Column 6 in Table~\ref{tab:1d}
shows the ratio of cold to hot mass accretion rates, $\chi$, computed
by averaging $\dot{M}'s$ over the radius and simulation time.
We average $\dot{M}'s$ over $r<100$ pc because the cadence of our data dumps is
comparable to the dynamical time scale at 100 pc. The larger the luminosity
the more matter is accreted via the cold phase. However, the maximum
value of $\chi$ is of the order of a few, so the dominance of the cold gas is
not too strong. 

\newpage
\begin{table*}[ht!]
\begin{center}
\begin{tabular}{ccccccccccc}
  \hline
  Model ID & $f_X$ & $N_r$ & $N_\theta$ & $t_f$ &
  $\langle\dot{M}\rangle$& $\langle f_{Vol}\rangle_{t}$ &$\langle\chi\rangle_{r,t}$&  $\langle\tau_{X,sc}\rangle_{\theta,t}$ & Max($\tau_{X,sc}$)&final state \\
  & &  & & [Myr] &${\rm [M_{\odot} \, yr^{-1}]}$ & & & & \\
  \hline
  2D256x64A  & $5 \times 10^{-4}$  & 256  & 64  & 20  & 2.0  &$1\times10^{-6}$&0& 0.45 & 0.46 &smooth\\
  \hline
  2D256x64B  & $8 \times 10^{-3}$ & 256  & 64  & 15.4 & 2.04 &$1\times10^{-6}$&$10^{-6}$& 0.1 & 0.99 &smooth\\
  2D512x128B & $8 \times 10^{-3}$ & 512  & 128 & 20   & 1.95
  &$1\times10^{-4}$& 0.02& 0.11& 4.7  &clouds  \\
  2D1024x256B& $8 \times 10^{-3}$ & 1024 & 256 & 1.83 & 1.84 &$1.5\times10^{-3}$&0.39& 0.19& 24.  &clouds\\
  \hline
  2D256x64C  & $1 \times 10^{-2}$ & 256  & 64  & 11.8 & 1.94 &$7\times 10{-5}$&0.15& 0.11 & 2.5 &smooth\\
  2D512x128C & $1 \times 10^{-2}$ & 512  & 128 & 20   & 1.88 &$5\times10^{-4}$&0.09& 0.13 & 17.3&clouds  \\
  2D1024x256C& $1 \times 10^{-2}$ & 1024 & 256 & 1.12 & 1.95 &$4\times10^{-3}$&0.5& 0.24 & 70 &clouds\\
  \hline
  2D256x64D  & $1.5 \times 10^{-2}$& 256 & 64   & 12   & 1.57 & $5\times10^{-3}$&0.3& 0.14 & 37.8 &clouds\\
  2D512x128D & $1.5 \times 10^{-2}$& 512 & 128  & 11   & 1.6  & $3\times10^{-3}$&0.43& 0.12 & 13.2 &outflow,filaments\\
\hline
\end{tabular}
\caption{
List of 2-D solutions. Columns from left to right are: model ID, $f_X$
- dimensionless luminosity, $N_r, N_{\theta}$ - numerical resolution
in radial and tangential direction, $t_f$ - final time of the
simulation, $\langle\dot{M}\rangle_{t}$ - time averaged mass accretion
rate, $\langle f_{Vol}\rangle_{t}$ - time averaged volume filling
factor, $\chi$ - a ratio of cold to hot mass accretion rate averaged
over radii less than $10^{20}$ cm and simulation time, (cold phase is
any gas with temperature $T<10^5$ K),
$\langle\tau_{X,sc}\rangle_{\theta,t}$ - angle and time averaged
optical depth, Max($\tau_{X,sc}$) - maximum optical depth recorded during the
simulation, and comments. In the beginning of all runs the accretion
flow becomes unsteady due to randomly seeded perturbations, at
$t_f$ the flow either returns to the original, unperturbed,
smooth state (globally stable solutions), or remains unsteady with
coexisting, two phase medium (cold clouds embedded in a warm
gas). Only in one model a large scale outflow forms (model
2D512x128D).}\label{tab:2d}
\end{center}
\end{table*}

\section{Results: 2-D models}\label{sec:results_2d}

\subsection{Seeding the TI}

To investigate the growth of instabilities in 2-D,
we solve eqs.~\ref{eq:mass}, \ref{eq:mom}, \ref{eq:energy}, for
the same parameters $M_{BH}$ and $f_X$ as in \S~\ref{sec:results_1d},
but on a 2-D, axisymmetric grid with $\theta$ angle changing from 0 to
$90 \deg$. We use three sets of numerical resolutions described in
\S~\ref{sec:num_setup}. To set initial conditions in axisymmetric
models, we copy the solutions found in 1-D models onto the 2-D grid. In
case of time-independent 1-D models, our starting point, is the data
from $t=t_{f}$. We checked, for example, the runs 2D256x64 A, B,
C, and D (which are 2-D version of models 1D256A, B, C, and, D) are
time-independent at all times as expected.  In case of higher
resolution models, for which the 1-D, steady state models do not
exist we adopt a quasi-stationary data from the 1-D run early
evolution (at $t$ of a fraction of a Myr), during which the flow is
already relaxed from its initial conditions but the TI
fluctuations are not yet developed. Models which are time varying in 1-D,
in 2-D develop dynamically evolving spherical shells, as expected, 
also indicating  that our numerical code keeps a symmetry in higher dimensions.

To break the symmetry in 2-D models, we
perturb the smooth solutions adopted as initial conditions. The
perturbation of a smooth flow is seeded everywhere and has a small
amplitude randomly chosen from a uniform distribution.  The new
density at each point is $\rho=\rho_0 (1+ Amp*rand)$, where $rand$ is a
random number $rand \in (-1,1)$ and maximum amplitude $Amp=10^{-3}$. 
To seed the isobaric, divergence free fluctuations other (than
$\rho$) hydrodynamical variables are left unchanged. The amplitude
magnitude $Amp$ is chosen to be much higher in comparison to
$\epsilon_{machine}$, in order to investigate the development and evolution 
of strongly non-linear TI on relatively short time scales, starting directly 
from a linear regime. The list of all perturbed, 2-D models is given
in Table~\ref{tab:2d}.

\subsection{Formation of Clouds, Filaments, \&  Rising Bubbles}

For luminosities $L_X < 0.015~L_{Edd}$, the 2-D models show similar properties
to the 1-D models. The gas is thermally and convectivelly unstable within the
computational domain, and we observe that very tiny fluctuations in an
initially smooth, spherically symmetric, accretion flow, grow first linearly
and then non-linearly. Since the symmetry is broken the cold phase of
accretion forms many small clouds.  For $L_X = 0.015~L_{Edd}$ or higher, the
cold clouds continue to accrete but in some regions a hot phase of the gas
starts to move outward.

\begin{figure*}
\includegraphics[angle=0,scale=0.17]{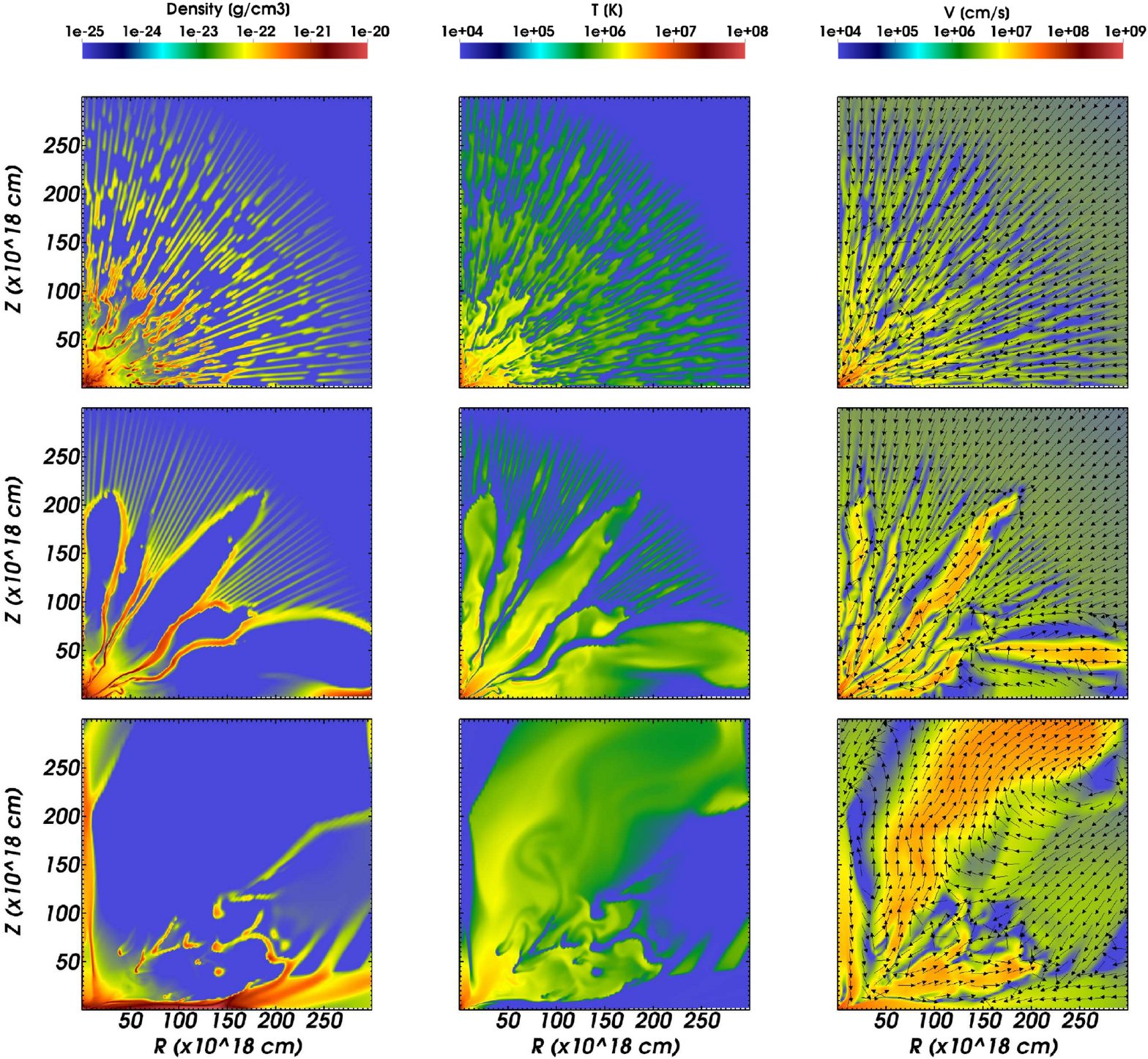}
\caption{
Density, temperature, velocity magnitude with velocity direction (left,
middle, and right column, respectively) in model 2D512x128D at t=3, 6 and 12
Myr (upper, middle, and bottom rows, respectively). Each panel show inner
parts of the flow at $r \le 100 $ pc.\label{fig:st2d}}
\end{figure*}

In Figure~\ref{fig:st2d}, we show three snapshots of representative 2-D model 
2D512x128D at various times (t = 3, 6 and 11.8 Myrs). This model has the best
resolution and the highest luminosity for which we are able to start the
evolution from nearly steady state conditions. Columns from left to right
show density, temperature, and total gas velocity overplotted with the
arrows indicating the direction of flow.  Initially (at t=3 Myr) the
smooth accretion flow fragments into many clouds, which are randomly
distributed in space. The cooler, denser regions are embedded in a
warm background medium. The colder clouds are stretched in the radial
direction and they have varying sizes. This initial phase of the
evolution is common for all models in Table~\ref{tab:2d}. 

The phase where many cold clouds accrete along with the warm background inflow
is transient. At a later stage (t=6 Myr, middle panels), model 2D512x128D
shows a systematic outflow in form of rising, hot bubbles. The outflow is
caused by the pressure imbalance between the cold and hot matter and buoyancy
forces. The hot bubbles expand at speeds of a few hundreds km/s. Despite of
the outflow, the accretion is still possible. During the rising bubble phase,
the smaller clouds merge and sink towards the inner boundary as 
streams/filaments.  However, even this phase is relatively short-lived.  Bottom
panels in Figure~\ref{fig:st2d} show the later phase of evolution when some of
the filaments occasionally break into many clouds (this process takes place
between 10 and 50 pc). These 'second generation' clouds occasionally flow out
together with a hot bubble. Along the X-axis, we see an inflow of a dense filament.

\begin{figure}[ht]
\includegraphics[angle=0,scale=0.4]{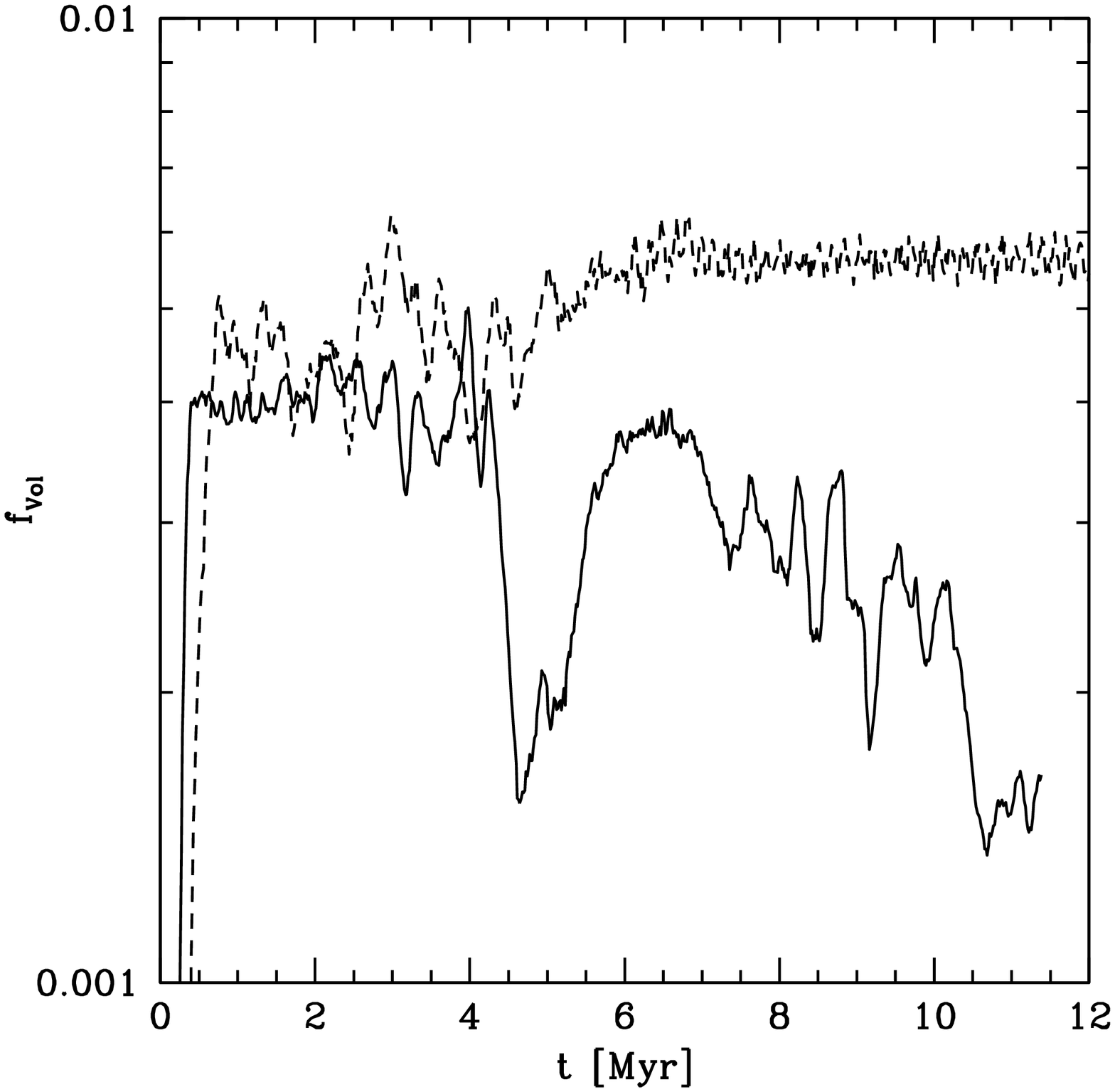}
\caption{Evolution of the volume filling factor $f_{vol}$ in model 2D512x128D
 (black, solid line) and 2D256x64D (dashed, blue line).
\label{fig:vol}}
\end{figure}

To quantify the properties of clumpy accretion flow, we measure the volume filling factor of a
cold gas $f_{vol}$, defined as:
\begin{equation}
f_{Vol}=V_{cloud}/V_{tot}
\end{equation}
where $V_{cloud}$ is the volume occupied by gas of $T<10^{5}$ K, and $V_{tot}$
is the total volume of the computational domain. In model 2D512x128D, the
time-averaged $f_{Vol}$ is $\langle f_{Vol}\rangle=3\times10^{-3}$. 
The time evolution of $f_{vol}$ within 60 pc is
shown in Figure~\ref{fig:vol} (black, solid line). The $f_{Vol}$ is variable
and at the moment of the outflow formation, $f_{Vol}$ suddenly decreases by a
factor of about 4. For comparison $f_{Vol}$ calculated during run 2D256x64D is
also shown (blue, dashed line). Run 2D256x64D has the same physical parameters
as 2D512x128D, however, no outflow forms.  In the latter case, $f_{vol}$ is
less variable and larger. In Table~\ref{tab:2d}, we gather the time averaged
$\langle f_{Vol}\rangle$ for all 2-D solutions. 
Measuring $f_{Vol}$ allows to quantify whether the perturbed 
accretion flow returns to its original, smooth state.
We find that this happens when the $\langle f_{Vol}\rangle\approx 10^{-5}$ or
smaller (models 2D256x64A, B, and C).

\subsection{$\dot{M}$ Evolution}

\begin{figure}[ht!]
\includegraphics[angle=0,scale=0.4]{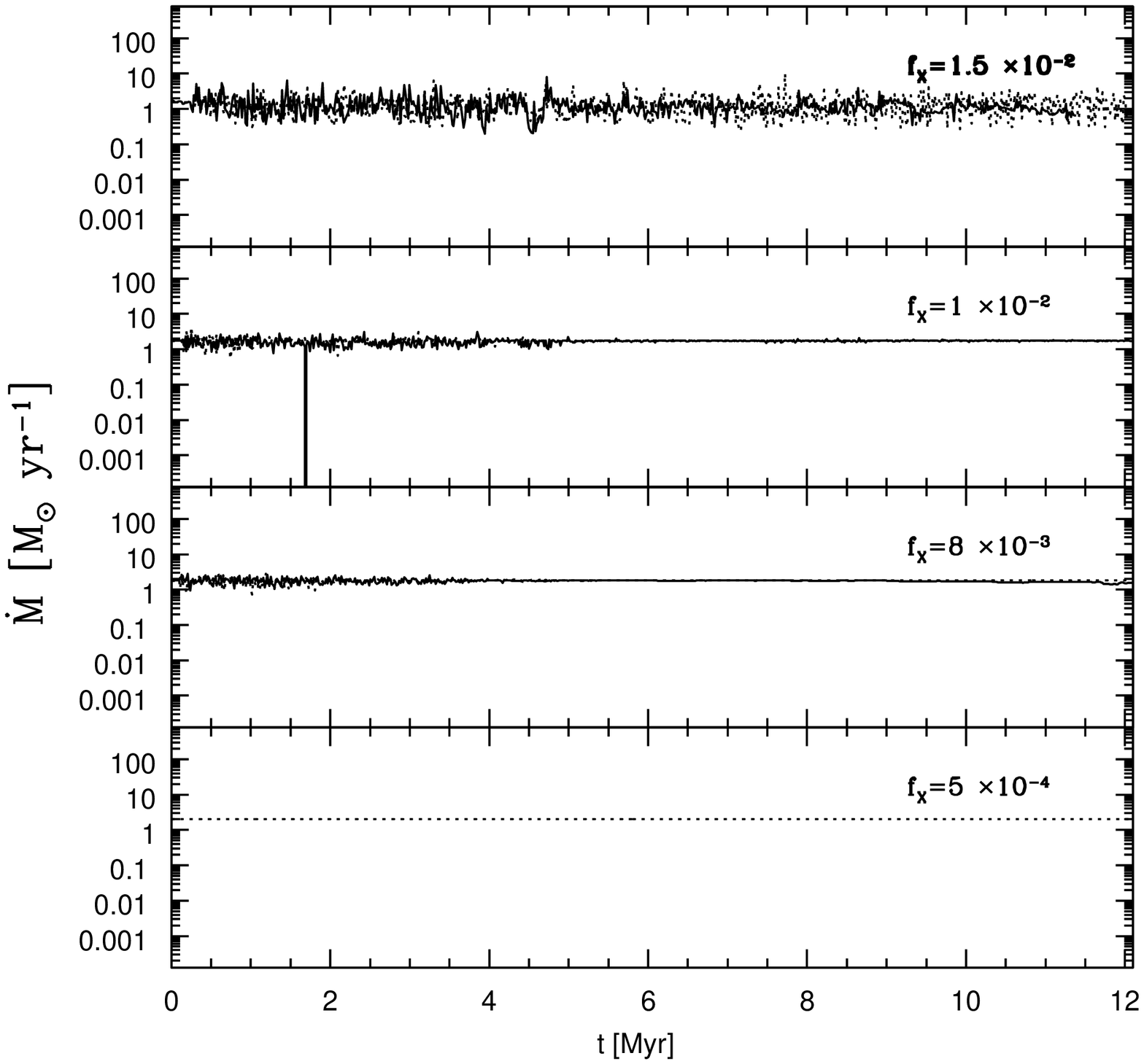}
\caption{Mass accretion rate in 2-D models with initially seeded
random perturbations.  Various colors code the $\dot{M}$ in models
calculated with various grid resolutions: 2D256x64 (blue), 512x128
(red), 1024x256 (green). Results are sensitive to the resolution same
as in 1-D models.~\label{fig:mdot2d}}
\end{figure}

Figure~\ref{fig:mdot2d} presents $\dot{M}$ through the inner boundary
measured as a function of time. In most cases (except model
2D256x64A), $\dot{M}$ becomes stochastic instantly with spikes
corresponding to the accretion of colder but denser clouds similar to
those in Figure~\ref{fig:st2d} (upper panels). Similar to 1-D models,
one can divide the solution into two types: steady and unsteady
state. In the latter, $\dot{M}$ fluctuates on various levels depending on $f_X$.

Table~\ref{tab:2d} lists several characteristics of our 2-D simulations, for
example, the ratio $\langle\chi\rangle_t$ averaged over time.
In 2-D models this variable is smaller in comparison to 1-D due to geometry of
the clouds. The maximum value of $\langle\chi\rangle_t$ is less than unity. 
This indicates that multi-dimensional effects (specifically development of
convection) promote hot phases accretions. We plan to investigate this issue
in future by carring out 3-D simulations. 

\begin{figure}[ht!]
\includegraphics[angle=0,scale=0.4]{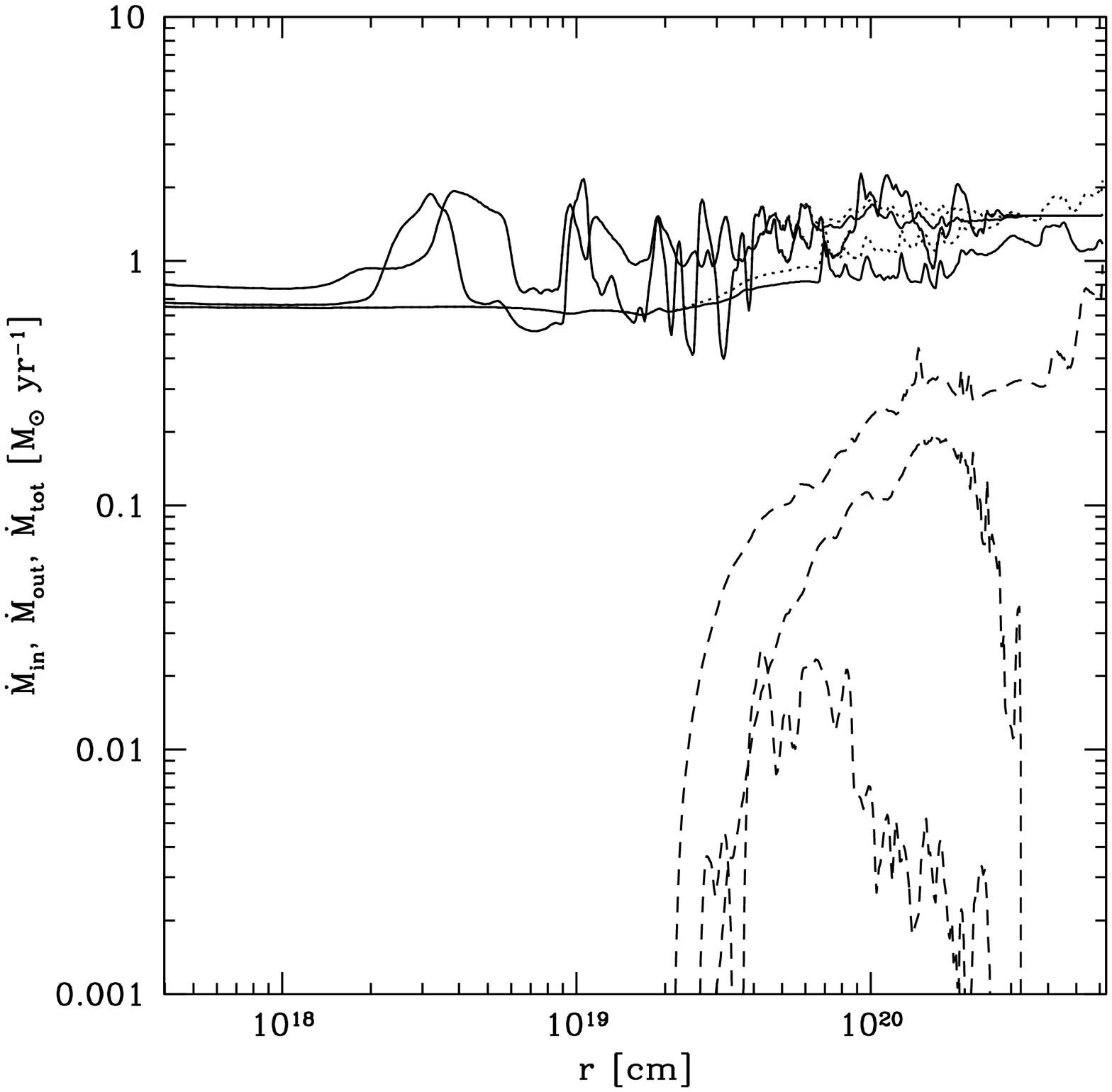}
\caption{Model 2D512x128D: 
In-, out- and total mass flow rate as a function of radius for three
time moments shown in Figure~\ref{fig:st2d} (green, blue, and black
correspond to t=3, 6 and 11.8 Myr). The solid lines mark the inflow rates
while the dashed line - outflow rate. The dotted line is the total
mass flow.~\label{fig:mdotinout}}
\end{figure}

We find that a large scale outflow forms only in run 2D512x128D. But
even in this case, the $\dot{M}$ is not significantly affected by the
outflow. Figure~\ref{fig:mdotinout} shows the mass outflow rate
(dashed line), inflow rate (dotted lines) and total mass flow rate
(solid line) as a function of radius. The same types of lines 
show $\dot{M}$ for
various times of the simulation (t=3, 6 and 11.8 Myr, green, blue and
black lines, respectively) and they are averaged over $\theta$
angle. The rising bubble originates at about 10 pc in this case.
We anticipate that large scale outflow are common and significant for high
luminosity cases (i.e., for $f_X > 0.02$).

\subsection{Obscuration Effects}

\begin{figure}[ht!]
\includegraphics[angle=0,scale=0.7]{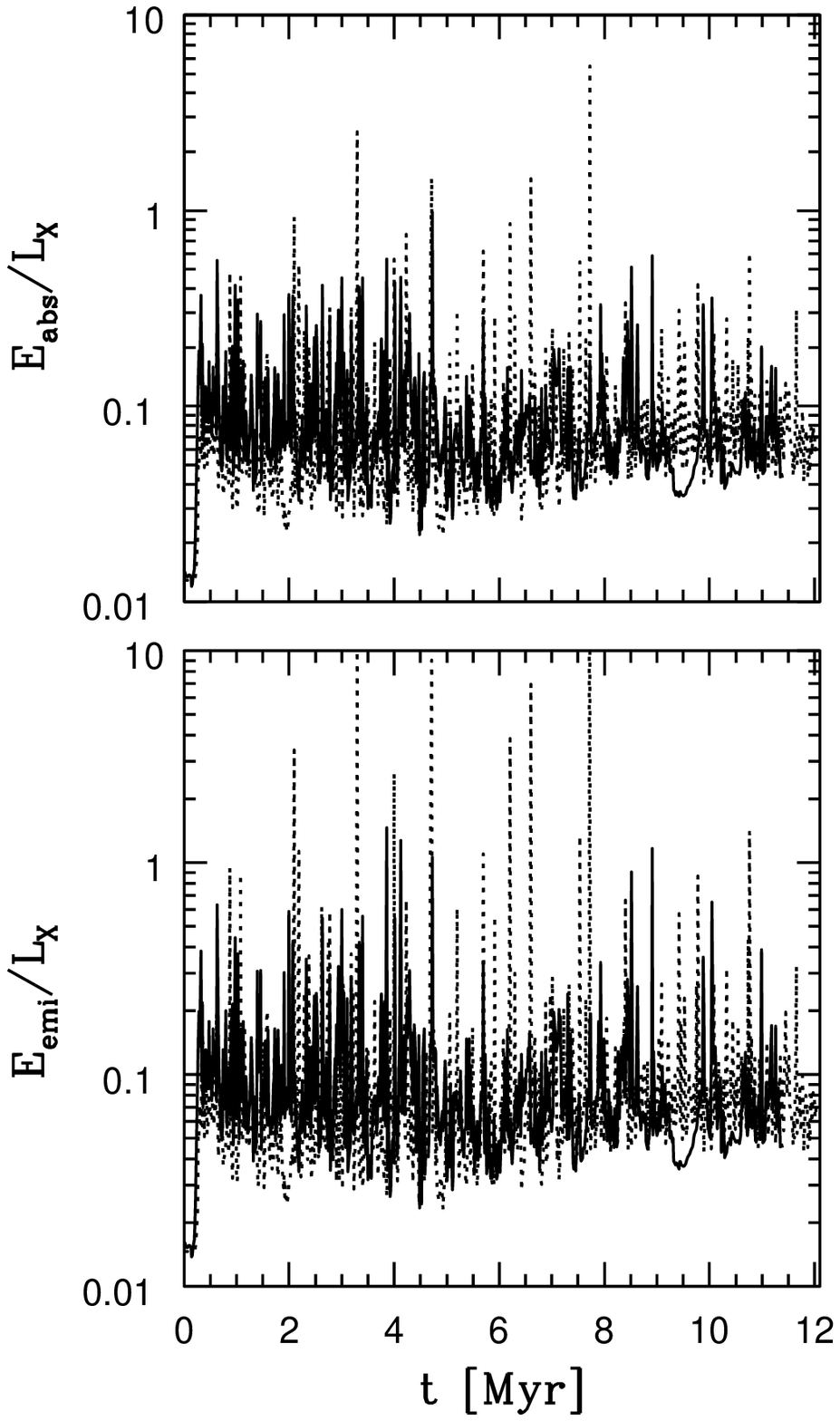}
\caption{Fraction of central illuminating source radiative energy
intrinsically absorbed (upper panel) and emitted (bottom panel) by gas per second as
a function of time in models 2D512x128 (red, solid line) and 2D256x64D (blue, dotted line). 
\label{fig:energy2d_all}}
\end{figure}

Here we again check whether the cloud opacity might affect our results.  The
averaged optical thickness of the filaments and clouds is similar (see
Table~\ref{tab:2d}, columns 8 and 9).  In Figure~\ref{fig:energy2d_all} we
show how much energy is absorbed and emitted in run 2D512x128D during the
evolution. The figure shows the intrinsic absorption and emission integrated over
the entire computational domain. We next calculate $\langle\tau_{X,cs}\rangle$
(optical thickness due to absorption, averaged over angles and times) and
maximum value of $\tau_{X,cs}$ that occurred during the evolution.  In case of
the largest optical depth of $\tau \approx 70$ (in run 2D1024x256C) the
radiation force coefficient from Equation~\ref{eq:rforce}: $f_{force} \approx
0.2$ which, as in 1-D models, is small but might not be negligible. Therefore,
we are planning to explore the effects of optical depth in a follow-up paper.

\section{Summary and Discussions}\label{sec:discussion}

In this work we show the evolution of thermal instabilities in gas accreting
onto a supermassive black hole in an AGN. A simplified assumptions made in
this work, in particular constant X-ray luminosity emitted near the central
SMBH regardless of the $\dot{M}$, allows to follow the development of TI from
the linear to strongly non-linear and dynamical stage up to luminosities of 
$L \approx 1.5 \times 10^{-2}~L_{Edd}$.

In our 1-D models the TI is seeded by numerical errors which
might be non-isobaric and are initially under-resolved. In the initial phase,
the TI growth rate is smaller than predicted by theory. The rate is affected by
grid resolution which leads to the formation of cold clouds of various
sizes and density contrasts. This is reflected in the mass accretion rate
fluctuating at different amplitude and rate for the same physical
conditions but different resolutions. One cannot avoid dealing with these numerical difficulties in the
numerical models. Nevertheless, we find the under-resolved, 1-D models very
useful in quick checking where the thermally unstable zone exists and what type of
fluctuation could cause the smooth to turn into a two-phase
medium. For given physical conditions, Figure~\ref{fig:res} shows the wavelength
$\lambda_0$ and Equation~\ref{eq:amp} gives the amplitude of an isobaric
perturbation required to break the smooth flow into a two-phase,
time-dependent model.

In 2-D models, although the models depend on the resolution effects same way
as in 1-D setup, we can observe an outflow formation. The convectivelly
unstable gas buoyantly rises and, as found in this work, controls the later
evolution of the two-phase medium and mass accretion rate. 
Given a simple set up with minimum number of processes included, our models display the
three major features needed to explain some of the AGN observations: cold inflow, hot
outflow and cold, dense clouds which occasionally escape, advected with the
hot wind.  We show that an accretion flow at late, non-linear stages, thus
most relevant to observations, are dominated by buoyancy instability not TI. 
This suggests that the numerical resolution might not have to be as high as that
needed to capture the small scale TI modes and it is sufficient to capture
significantly larger and slower buoyancy modes. 
We plan to check the consistency of the models with the observations by
  calculating the synthetic spectra, including emission and absorption lines
  based on our simulation following an approach like the one in
  \citet{sim:2012}.  Here, we only briefly comment on the main outflow
  properties and compare them some observations of outflows in 
  Seyfert galaxies.

Space Telescope Imaging Spectrograph (STIS) on board 
the {\it Hubble Space Telescope}
allows us to map the kinematics of the Narrow Line Regions in some nearby 
Seyfert Galaxies (e.g. for NGC 4151 \citealt{das:2005};
NGC 1068; \citealt{das:2006}; Mrk 3, \citealt{crenshaw:2010}; 
Mrk 573, \citealt{fisher:2010}; and Mrk 78, \citealt{fisher:2011}).
Position-dependent spectra in [O III] $\lambda$~5007 
and $H_{\alpha}$ $\lambda$~6563, and the measurements
of the outflow velocity profiles show the following general trend:
the outflow has a conical geometry and the 
[O III] emitting gas accelerates linearly up to some radius 
and then decelerates. The velocities typically reach up to about 1000 km/s and 
a turnover radius is on one hundred to a few hundred parsec scales. 

\begin{figure}[!hr]
\includegraphics[angle=0,scale=0.4]{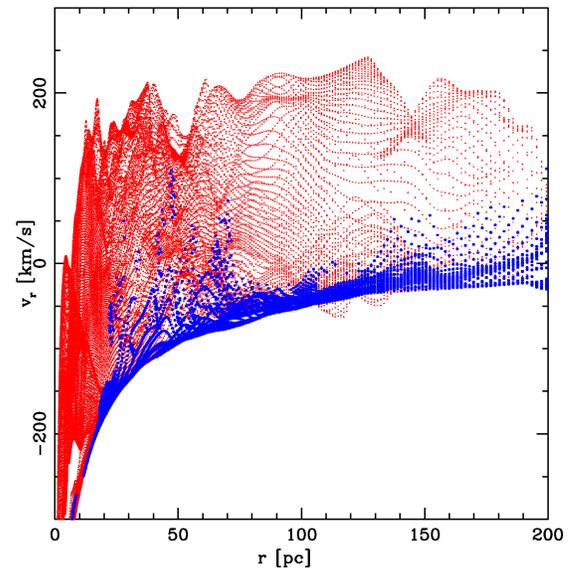}
\caption{
Scatter plot of radial velocity of hot ($T>10^5 K$, smaller red symbols) and
cold ($T < 10^5 K$, larger blue symbols)
phase of the flow in model 2D512x128D at t=11.8 Myr (model shown in right
panels in Figure~\ref{fig:st2d}).}
\label{fig:vr_scatter}
\end{figure}

To compare our results with the observations, Figure~\ref{fig:vr_scatter} 
shows the radial velocity of hot and
cold gas versus the radius at t=11.8~Myr for model 2D512x128D
(the data correspond to a snapshot shown in the right panels 
in Figure~\ref{fig:st2d}). We reiterate that our model is 
quite simplified (e.g., no gas rotation) and the outer radius is relatively 
small (i.e., 200~pc). Therefore, our comparison is only
illustrative.

We find that
the hot outflow originates at around 10 pc and accelerates up to about
$v_{max} \approx 200 {\rm km/s}$, which is comparable to the escape velocity 
from 10~pc, $v_{esc}=314~{\rm km/s}$. At larger distances, r = 100-200~pc, 
we see a signature of deceleration which is consistent with 
the observations of Seyferts outflows. 
We note that the geometry of the simulated flow is affected by 
our treatment of the boundaries of the computational domain, specifically
along the pole and the equator we use
reflection boundary conditions (see \S~\ref{sec:num_setup}).

The scatter plot also indicates that the cold clouds appear at about 20-80 pc.
Their maximum velocity is about $v_{max}=100$ km/s, which is smaller than
the velocity of the hot outflow. The plot does not show a clear
indication of a linear acceleration of the outflowing cold gas.
However, it is possible that the cold clouds, seen in this snapshot,
will continue to be dragged by the hot outflow and eventually
will reach higher velocities.

We also measured the column density of the hot and cold gas for the same,
representative, snapshot at t=11.8 Myrs. 
The typical column densities vary with the observer inclination, 
$N_{H}=5 \times 10^{22} - 10^{24} \, {\rm cm^2}$ for gas $T >
10^5$ K, and $N_{H}=10^{20}-10^{23} \, {\rm cm^2}$ for gas with $T<10^5$ K. 
This is roughly consistent with column densities estimated from 
observations of AGN (e.g. for NGC 1068 $N_H= 10^{19}-10^{21} \, {\rm cm^2}$,
\citealt{das:2007} and references therein).

Our results are similar in many respects, to the previous findings presented
in \citet{barai:2012}, i.e.  the accretion evolution depends on $f_X$
luminosity; we also observe clouds, filaments and outflow. The outflow appears
at $f_X=0.015$ which is consistent with $f_X=0.02$ found by
\citet{barai:2012}.  Here, we are able to calculate models for about 10 times
longer in comparison to 3-D SPH models. We confirm the previous results that
the cold phase of accretion rate can be only a few times larger in comparison
to the hot one.

Similar models has been investigated in the past by
e.g. \citet{krolik:1983}. Our work is on one hand a simplified and on the
other hand an extended version of these previous works. The key extention here
is that our new results cover the non-linear phase of the evolution. There are
two new conclusions added by our analysis to the previous investigations.  The
2-D models with outflows are possibly governed by other than TI instabilities
mainly convection. Another non-linear effects found in our 1-D and 2-D models is
that the fragmentation of the flow makes it optically thick for photoionization.
Further investigation of shadowing effects is required.

Some sub-resolution models of AGN feedback in galaxy formation (\citealt{dimatteo:2008};
\citealt{dubois:2010}; \citealt{lusso:2011}) assume that BH accretion is
dominated by an unresolved cold phase, in order to boost up the accretion rate obtained
in simulations. Our results indicate that the cold phase
accretion is unlikely dominant as even in well-developed and well- resolved
multi-phase cases, the accretion is typically dominated by a hot phase.
However, we note that the cold phase of our solution might be an upper branch
of some more complicated multi-phase medium (i.e., a mixture of molecular,
atomic and dusty gas).

This work was intentionally focused on a very limited number of processes and
effects. Its results suggest that the future work should include more
self-consistent approach not only with shadowing effects but also with the
radiation force. Our next step would be to investigate the non-axisymmetric
effects via fully 3-D simulations.  The latter is challenging and one may not
be able to see very fine details of the gas dynamics as in 2-D models due to
resolution effects.

\acknowledgements
This work was supported by NASA under ATP grant NNX11AI96G and NNX11AF49G.  DP
thanks J. Ostriker, J. Stone, and S. Balbus for discussions and also
Department of Astrophysical Sciences, Princeton University for its hospitality
during his sabbatical.  DP also acknowledges the UNLV sabbatical assistance.
Authors would like to thank Paramita Barai, Ken Nagamine and Ryuichi Kurosawa for
their comments on the manuscript.

%\bibliographystyle{apj}
%\bibliography{local}

\appendix

\section{Growth rate of a condensation mode in a uniform medium -code tests}\label{app1}

\citet{field:1965} formulated a linear stability analysis of a gas in thermal
and dynamical equilibrium. Here, we briefly recall his most important, for our analysis, 
equations. We disregard the thermal conduction
effects. The dispersion relation derived from linearized local fluid equations
with heating/cooling described by ${\mathcal L}$ function and perturbed by a periodic, small amplitude
wave given by $\exp(nt+ikx)$, is:
\begin{equation}
n^3 + N_v n^2 + k^2 c_s^2 n + N_p k^2 c_s^2 = 0 \label{eq:cube}
\end{equation}
where $k$ is the perturbation wave number ($k=2 \pi /\lambda$) and functions $N_p$ and $N_v$ are defined as
\begin{equation}
N_p \equiv \left .\frac{1}{c_p} \left(\frac{\partial {\mathcal L}
}{\partial T}\right)\right|_P \label{eq:Np}
\end{equation}
and
\begin{equation}
N_v \equiv \left. \frac{1}{c_v} \left(\frac{\partial {\mathcal
    L}}{\partial T}\right)\right|_\rho \label{eq:Nv}
\end{equation}
with $c_p$ and $c_v$ being the specific heats under constant pressure and
constant volume conditions, respectively, and $T$ is the gas
temperature. Vertical line means that the derivative is taken under constant
thermodynamical variable condition. Dispersion Equation~\ref{eq:cube} has
three roots. In a short wavelength regime ($\lambda \ll 2\pi N_p / c_s$), two,
complex roots correspond to two conjunct nearly adiabatic sound waves and
third, real one is an isobaric condensation mode (the gas density and
temperature change in anti-phase so that the pressure remains constant).  The
sign of the real part of the root gives the stability criterion. The sound
wave will grow if $\left. \partial{\mathcal L}/\partial T \right|_S < 0$
(known as Parker's criterion, \citealt{parker:1953}).  The condensation mode
will grow if $\left. \partial {\mathcal L}/\partial T \right|_P < 0$ (Field's
criterion). In a short wavelength limit, the growth rates asymptote to
$n=-0.5(N_v-N_p)$ (for sound waves) and $n=-N_p$ (for condensation
modes). Isochoric modes ($n \rightarrow -N_v$) and effective acoustic waves
are eigen modes of long wavelengths perturbations. The perturbation
growth/damp time scale is $\tau_{TI}=1/n$.

We use the above \citet{field:1965} theory to show that our numerical scheme
for solving the modified energy conservation equation
(Equation~\ref{eq:energy}) together with two other fluid dynamics equations is
accurate. The test calculations are carried out in 1-D Cartesian coordinates
within $x\in(0,L)$ range where L is the size of the computational domain in
dimensionless units. The boundary conditions for all variables are periodic.
In an unperturbed state, the gas density ($\rho_0=1$) and internal energy
density ($e_0=1$) are constant in the entire computational domain.  The
velocity of gas is set to zero. We assume that the gas is heated by an
external source of radiation and cools due to free-free transitions. The test
cooling function is simple:
\begin{equation}
{\mathcal L}= C \rho T^{1/2}  \label{testcool} - H 
\end{equation}
The normalization constants
$H$ (for heating) and $C$ (for cooling) are set so that in the unperturbed state
the gas is in radiative equilibrium i.e. ${\mathcal L}(\rho_0,e_0)=0$.
In this test the functions $N_p$ and $N_v$ have explicit, analytical forms
\begin{equation}
N_p \equiv \left .\frac{1}{c_p} \left(\frac{\partial {\mathcal L}
}{\partial T}\right)\right|_P \equiv
  \frac{1}{c_p} \left( \left . \frac{\partial{\mathcal L}}{\partial T}
  \right |_\rho -
 \frac{\rho}{T} \left . \frac{\partial{\mathcal L}}{\partial \rho}
\right |_T \right)
=- \frac{1}{2 c_p} C \rho_0 T_0^{-1/2}
\end{equation}
and
\begin{equation}
N_v \equiv \left. \frac{1}{c_v} \left(\frac{\partial {\mathcal L}}{\partial T}\right)\right|_\rho
= \frac{1}{2 c_v} C \rho_0 T_0^{-1/2}  = -\gamma N_p.
\end{equation}
The numerical values of limiting growth/damp rates are $N_p=-0.04$,
and $N_v=0.067$, while the speed of sound is: $c_s^2=1.11$
($\gamma=5/3$). The domain sound crossing time is much shorter than
the perturbation growth time scale which allows to keep the constant pressure.
%$N_p=-0.0404698666921181477$,
%$N_v=0.06744975209$. 
%$n=0.5(N_p-N_v)=-0.053960$. 
%$c_s^2=1.1111$. 

\begin{figure*}[ht]
\includegraphics[angle=0,scale=0.4]{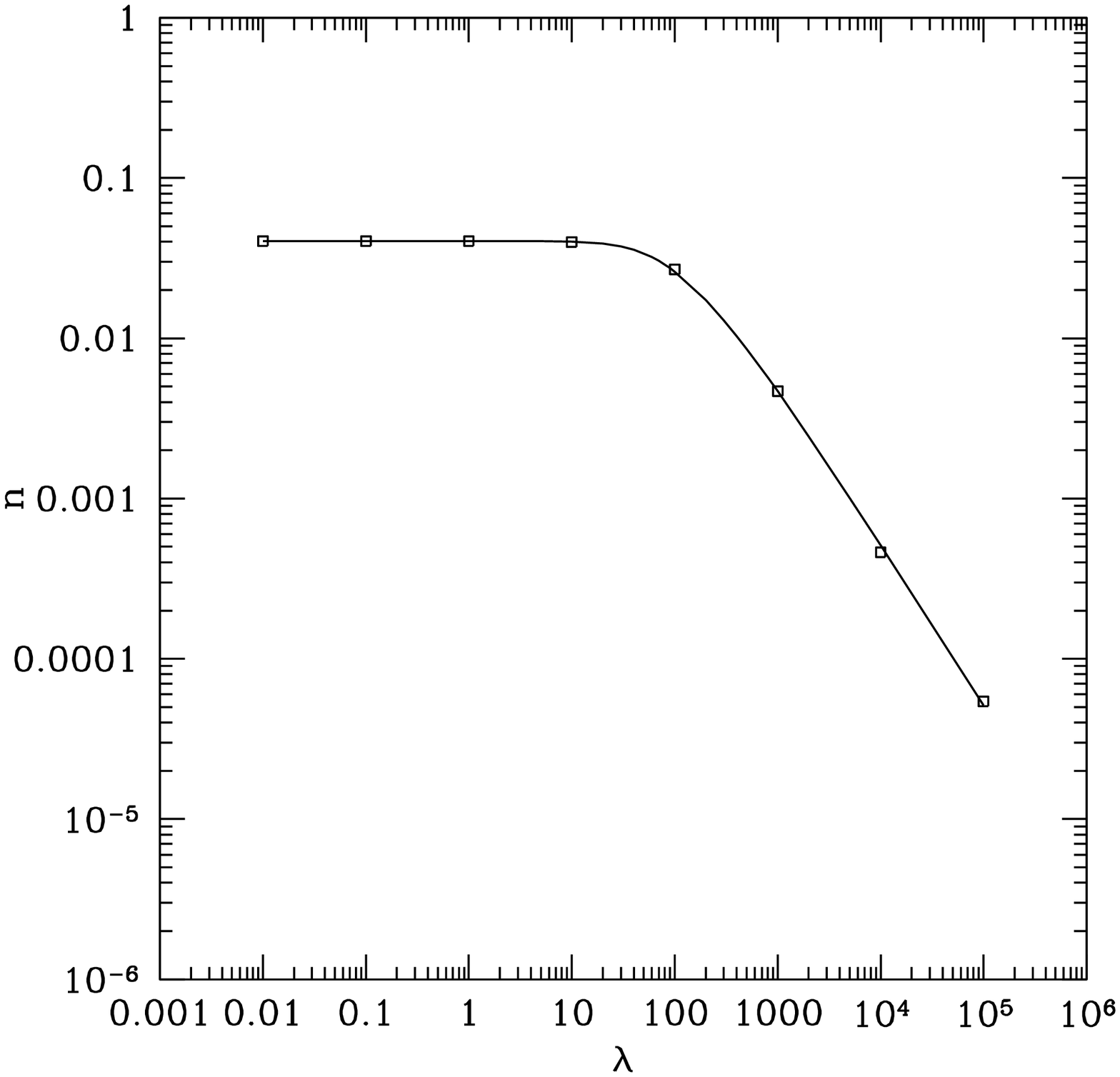}
\includegraphics[angle=0,scale=0.4]{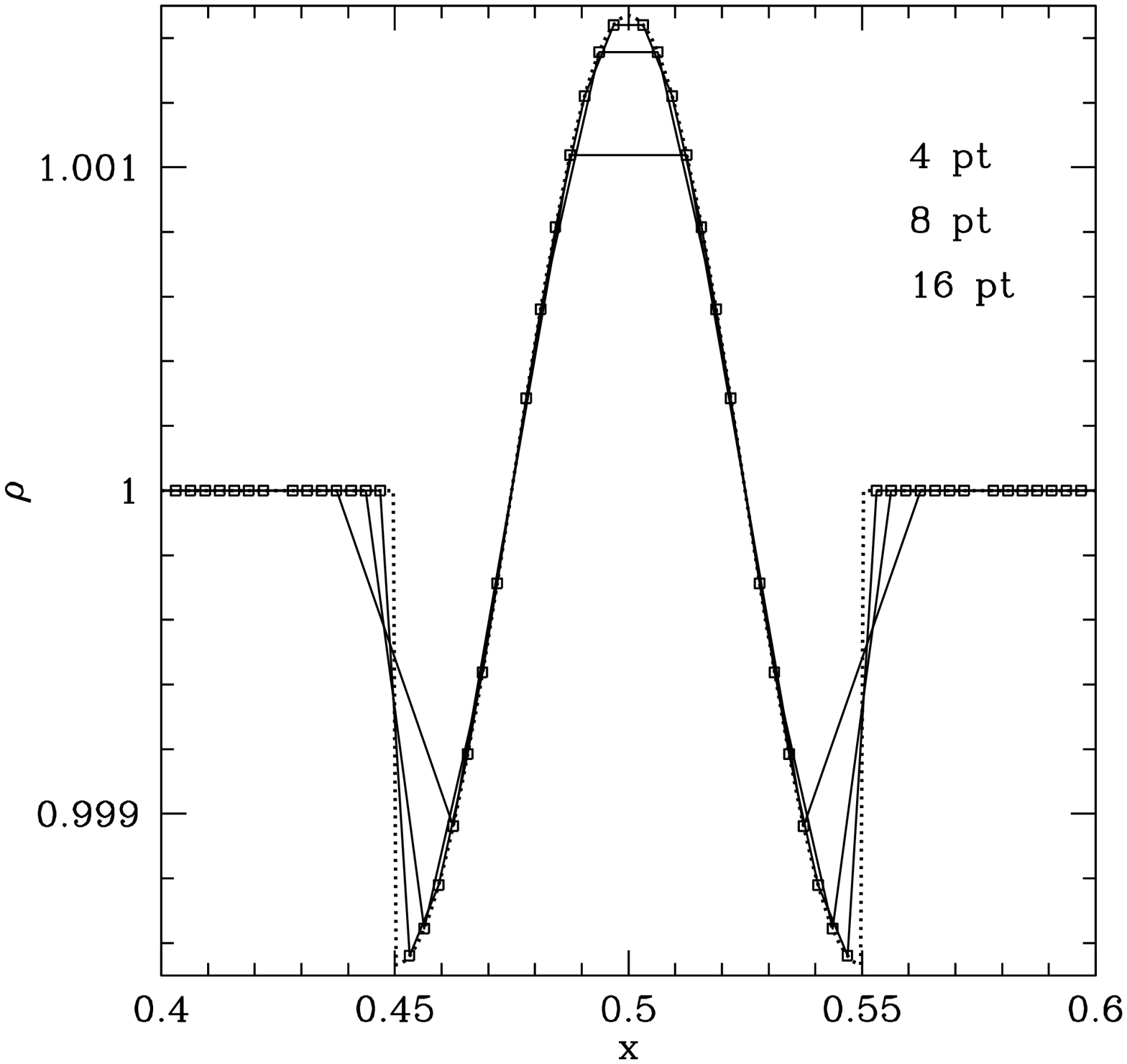}
\caption{
Left panel: Analytical (solid line; third root of Equation~\ref{eq:cube}) and
numerical (points; calculated with ZEUS-MP) linear growth rates of an eigen
mode with wavelength $\lambda$. The initial amplitude of a perturbation is
$A=10^{-3}$. The numerical solution uses $N_x=64$ grid points.  Right panel:
The condensation mode (density profile) with $\lambda=0.1L$ shown at t=10 T
(where T here is a sound crossing time over the computational domain $L=1$)
when it is resolved by 4, 8 and 16 points. The dotted line shows the very high
resolution of 1600 points for which the condensation mode grows at the
expected theoretical rate. Here the growth rate is as expected when resolved
by at least with 16 points. For lower resolutions the condensations grow
slower than one would expect.
\label{fig:app}}
\end{figure*}

Our numerical scheme implemented into ZEUS-MP code
correctly reproduces the expected growth rates of small amplitude perturbation 
of the uniform medium. 
The perturbation is an eigen mode of TI, and its properties depend on the
assumed $\lambda$. Eigen modes are realized by first applying a cosine
perturbation to the gas density $\rho= \rho_0 + A \rho_0 \cos(k x)$ 
and calculating profiles of $e$ and $v$ from
e.g. Equations 11 and 14 in \citep{field:1965}, for a given $k$ and
corresponding theoretical value of $n$ (given by Equation~\ref{eq:cube}).
Next we measure how fast the perturbation grows while it is in the linear
regime. Figure~\ref{fig:app} (left panel), shows the analytical solution of
the theoretical dispersion relation $n(\lambda)$ (solid line, third root of
Equation~\ref{eq:cube}), and the numerical growth
rates calculated with ZEUS-MP (points). For very short $\lambda$'s the eigen
mode of this root is converging to the isobaric condensation mode and grows at
$n=-N_p$ rate, as expected. The long $\lambda$ modes grow slower in comparison
to the very short $\lambda$ condensations, as predicted by theory. For
relatively large  $\lambda$, the third root changes into an effective acoustic
wave, it becomes complex with the real part negative meaning that the waves
are damped (see \citealt{shu:1992}, Equation 41 in the Problem Set No 3).

In the second test, we measure the growth rate of a condensation mode that
has a finite size (i.e. smaller than the domain length). We are interested in
how many numerical grid points is required to resolve the correct $n$. We set
$\lambda=0.1$ while $L=1$. Figure~\ref{fig:app} (right panel) shows
the same time snapshots of the growing condensation mode density, calculated with
various numerical resolutions. Models with lower resolution evolve
slower. When $\lambda$ resolved with 16 points it starts converging to the
right solution. We conclude that about 20 or more grid points per $\lambda$ is
required to resolve the isobaric condensation.

\end{document}